\documentstyle[12pt,epsfig,amsfonts,latexsym]{article}

\makeatletter

\@addtoreset{equation}{section} \makeatother

\newcommand{\be}{\begin{equation}}
\newcommand{\ee}{\end{equation}}
\newcommand{\ba}{\begin{eqnarray}}
\newcommand{\ea}{\end{eqnarray}}

\newcommand{\dle}[1]{\label{#1}}
\newcommand{\dr}[1]{\ref{#1}}
\newcommand{\dc}[1]{\cite{#1}}
\newcommand{\dcom}[1]{}
\newcommand{\dnote}[1]{}

\newcommand{\nn}{\nonumber}

\newcommand{\gsim}{\raise.3ex\hbox{$>$\kern-.75em\lower1ex\hbox{$\sim$}}}
\newcommand{\lsim}{\raise.3ex\hbox{$<$\kern-.75em\lower1ex\hbox{$\sim$}}}

\begin{document}

\renewcommand{\thefootnote}{\fnsymbol{footnote}}

%-------------------------------------------------

%\titlepage
\begin{flushright}
LPT-ORSAY 03/31 
\end{flushright}
\vskip 1cm
\begin{center}
{\Large \bf Tachyon kinks on non BPS D-branes }
\end{center}
\vskip 1cm
\begin{center}
Ph.Brax $^{a}$\footnote{{\tt
brax@spht.saclay.cea.fr}}, J.Mourad $^{b}$\footnote{{\tt
mourad@th.u-psud.fr}} and D.A.Steer $^{b}$\footnote{{\tt
steer@th.u-psud.fr}}
\\
\vskip 5pt \vskip 3pt {\it a}) Service de Physique Th\'eorique,
CEA Saclay, 91191
Gif-sur-Yvette, France
\vskip 3pt
{\it b})
Laboratoire
de Physique
Th\'eorique\footnote{Unit\'e Mixte de Recherche du CNRS (UMR 8627).}, B\^at. 210, Universit\'e
Paris XI, \\ 91405 Orsay Cedex, France\\
and
\\
F\'ed\'eration de recherche APC, Universit\'e Paris VII,\\
2 place Jussieu - 75251 Paris Cedex 05, France.
\\
\end{center}
\vskip 2.0cm
\renewcommand{\thefootnote}{\arabic{footnote}}
\setcounter{footnote}{0} \typeout{--- Main Text Start ---}

\begin{center}
{\large Abstract}
\end{center}
\noindent

We consider solitonic solutions of the DBI tachyon effective action
for a non-BPS brane. When wrapped on a circle, these solutions are
regular and have a finite energy. We show that in the
decompactified limit, these solitons give Sen's infinitely thin
finite energy kink --- interpreted as a BPS brane --- 
provided that some conditions on the potential hold. In
particular, if for large $T$ the potential is exponential, $V =
e^{-T^a}$, then Sen's solution is only found for $a<1$.  For
power-law potentials $V = 1/T^b$, one must have $b > 1$. If these
conditions are not satisfied, we show that the lowest energy
configuration is the unstable tachyon vacuum with no kinks.
We examine the stability of the solitons and the spectrum of 
small perturbations.

\newpage

\section{Introduction}

In addition to the stable and charged BPS D-branes
\cite{polchinski},
type II superstrings contain unstable and uncharged
non-supersymmetric D-branes \cite{sen}. The instability
of these non-BPS branes is signalled by the presence
of a tachyon on their worldvolume. The decay of 
non-BPS branes is described by the dynamics of this tachyon.
Sen has also proposed the remarkable fact that
 the BPS branes can also be viewed as tachyon kinks on the
 non-BPS branes with one dimension higher \cite{sen2}.
 Different approaches \cite{Sen:1999md,Garousi:2000tr,
 Garousi:2002wq,Bergshoeff:2000dq,Kluson:2000iy,
 Sen:2002an,Sen:2002qa,Kutasov:2003er,Garousi:2003pv}
 converged to the following
 Dirac-Born-Infeld (DBI) effective action describing the
 dynamics of the tachyon field $T$:
 \be
 S = - \int d^px \, dt \, V(T)
 \sqrt{\left( 1 +  \partial_\mu T \partial_\nu
T \eta^{\mu \nu} \right)}.
\dle{dbi}
\ee
Here $V(T)$ is the tachyon potential which is even and 
vanishes at infinity where it reaches its minimum.  Near the
global maximum at $T=0$, $V(T)={\cal T}_p(1-\beta^2 T^2/2)+\dots$
where ${\cal T}_p$ is the tension of the non-BPS $p$-brane and the
potenial encodes the mass of the tachyon near the perturbative
vacuum $T=0$.  Finally $\eta^{\mu \nu} = (-1,+1,\ldots,+1)$ is the
$p+1$ Minkowski metric. 

Finite energy solitons obtained from the DBI action and depending on one
spatial coordinate (kinks) are singular\footnote{see also section 2.}
\cite{Sen:2003tm, Minahan:2000tg,Lambert:2002hk}. 
The goal of the
present paper is to construct these kinks as limits of regular
solitons by compactifying the spatial coordinate and looking for
conditions on the potential which guarantee the existence of the
decompactified limit. We shall show that the existence of this
limit implies two new conditions on the potential $V(T)$ as $¦T¦
\rightarrow \infty$: {\it i)} that $|V'|/V$ must tend
to zero, and {\it ii)} that $|V'|/V^2$ must tend to infinity.

 The action (\ref{dbi})
is supposed to be valid for large $T$
and small higher order derivatives of $T$, and this is compatible
with the limit we are considering.
Our results, together with
the ones coming from time dependent solutions \cite{Sen:2002an,
Kutasov:2003er},
boundary string field theory \cite{bou}
and perturbative amplitude calculations \cite{Garousi:2003pv},
give constraints on the potential
of the tachyon in different regimes and hopefully may help to
determine an effective action capturing a large domain of
validity.  On the other hand, if the potential is determined by other means
and found not to satisfy the conditions {\it i)} and {\it ii)} 
above (as is the case for a number of potentials studied in the literature),
then we would conclude that the DBI has no stable kink solution.

The plan of this paper is as follows.
In Section 2 we revisit Derrick's theorem which gives necessary
conditions for the existence of finite energy solitons
for actions of the form (\ref{dbi}). Within this framework we show how
the infinitly thin kink may appear.
In Section 3, we compactify one spatial coordinate
on a circle of radius $R$ in order to
get regular kinks.
In fact, we obtain regular solitons describing
$n$ pairs of kinks and anti-kinks. This is to be expected since the
net RR charge is zero in a compact space.
The kink and anti-kink are separated by a distance
$2\zeta=2\pi R/2n$ and the energy of the configuration
has the form $2n{\cal E}(\zeta)$. The decompactified limit
corresponds to $\zeta\rightarrow \infty$ with $\cal E$ finite
in the limit. In Section 3.1 we determine the conditions
on the potential, stated above,
for which this limit exists. We also give some
examples; in particular, for exponentially decaying potentials
of the form $e^{- T^a}$ we show that the conditions are
satisfied if and only if $a<1$.  For power-law decay, $V=T^{-b}$,
they are satisfied if and only if $b>1$.
In Section 4 we examine the
stability of the solitons and show that the unstable fluctuations
disappear in the decompactified limit whenever this limit is well
defined.  Finally conclusions are presented in Section 5.

\section{Solitons and Derrick's theorem}

We aim to search for solitonic or kink-like solutions of
(\dr{dbi}). These kink solutions should represent the stable BPS
$p-1$ branes into which the non-BPS $p$-brane decays.  Before
doing this explicitly, it is worthwhile recalling Derrick's
theorem \cite{Rajaraman:is}: in the case of the usual Klein-Gordon
action for scalar fields, it tells us that finite energy static
solitonic solutions on an infinite space are only possible in
(1+1)-dimensions. In this section we will draw similar conclusions
starting from (\dr{dbi}).

In fact, to make contact with the literature, we will initially
consider a slightly more general action:
\be
S = - \int d^px \, dt \, V(T) \left( 1 +
\partial_\mu T \partial_\nu T \eta^{\mu \nu} \right)^q,
\dle{actiongen}
\ee
where the real scalar field $T$ is dimensionless (we set
$\alpha'=1$ throughout). For $q=1/2$ this is just (\dr{dbi}). The
case $q=1$ has also been proposed as an effective tachyon action
(see, for instance, \dc{Lambert:2002hk}), whose kinks were studied
in \dc{hash}.  We will recover some properties of these kinks using
Derrick's theorem.\footnote{Note that when $q=1$, a change of
variables $T=T(\phi)$ enables action (\dr{actiongen}) to be
written as a Klein-Gordon action for $\phi$ with potential
$W(\phi)=V(T(\phi))$. (However, $\phi$ is typically compact:
$|\phi| < K = \int_0^{\infty} \sqrt{V(T)} dT$ which is finite for
e.g.\  exponentially decaying potentials. Note also that $W(K)=0$). When
$q=1/2$ a similar field redefinition alone can no longer put
action (\dr{actiongen}) into the canonical KG form. However,
%as when going from the NG to the Polyakov action,
if the square root is first linearised by introducing an auxially
field $h_{\mu \nu}$ (c.f.\ the Nambu-Goto versus Polyakov
actions), then (\dr{actiongen}) can be written as the action for a
scalar field $\phi$ evolving in a gravitational field $h_{\mu
\nu}$  with a position-dependent potential $W(\phi) =V(T(\phi)) [ h^{\mu \nu}
\eta_{\mu \nu}  - (p-1)]$. In this form any intuition 
regarding solitonic solutions breaks down.}

To recall Derrick's theorem, consider first a canonical scalar
field $\phi(\vec{x},t)$ with action
\be
S_{\rm KG} = \int d^p x \, dt \left(- \frac{1}{2} \partial_\mu {\bf \phi}
\partial^\mu {\bf \phi} - W(\phi) \right)
\ee
where the $p$ spatial directions are infinite, and the potential
$W(\phi)$ is non-negative vanishing only at its absolute minima.
Let $\phi_1(x)$ be a static solution of the equations of motion,
and hence an extremum, $\delta E=0$, of the finite static energy
functional
\be
E[\phi] = \int d^p x \left( \frac{1}{2} \partial_i {\phi}
\partial^i {\phi} + W(\phi) \right) \qquad (i=1,\ldots,p).
\ee
Now consider a one parameter family of solutions
$ \phi_{\lambda}(x) = \phi_1(\lambda x)$.
Since $\phi_1(x)$ is an extremum of $E$, then
$\left.\frac{d}{d\lambda} E[\phi_\lambda]  \right|_{\lambda = 1} = 0$
leading to
\be
0 = \int d^p x \left(p \, W(\phi) + (p-2) \frac{1}{2} \partial_i
{\phi}
 \partial^i {\phi} \right).
\dle{constraintphi}
\ee
Thus when $p \geq 2$, only the vacuum solution ($W(\phi)=0$,
$\phi=$const) is allowed.  Only if $p=1$ may non-trivial
${\vec{x}}$-dependent solutions exist.

We now carry out a similar analysis for the tachyon action
(\dr{actiongen}). The static energy functional is
\be
E[T] = \int d^p x V(T)  \left( 1 + \partial_i T \, \partial^i T
\right)^q,
\dle{Etach}
\ee
and proceeding as above leads to the constraint
\be
0=\int d^p x V(T) \left( 1 + \partial_i T \, \partial^i T
\right)^{q-1} \left[  p+ (p-2q)  \partial_i T \,\partial^i T \right] .
\dle{derrickT}
\ee
Notice that, as opposed to (\dr{constraintphi}), the potential $V(T)$
appears as an overall multiplicative factor.

When $q=1$ the term in square brackets in (\dr{derrickT}) only
vanishes and gives non-trivial solutions if $p=1$, {\it i.e.}\ in
(1+1) dimensions.  In this case, (\dr{derrickT}) imposes that
\be
T'^2 = 1 \qquad \Longrightarrow \qquad T = \pm x.
\dle{kink1}
\ee
(From now on, the kink will be taken to be centered about $x=0$:
$T(0)=0$.)
 This kink/anti-kink solution has been discussed elsewhere
\cite{hash} and it is easy to verify that $T=\pm x$ is indeed a
solution of the equations of motion.  Such solutions are
topological, interpolating from one vacuum to another, and from
(\dr{Etach}) their energy (or tension ${\cal T}_{p-1}$) is given
by
$$
E = {\cal T}_{p-1} =  2 \int_{-\infty}^{\infty} V(x) dx
$$
which, if $V$ decays sufficiently fast at infinity, is finite.
Notice that from the properties of $V(T)$, the energy density of
the field is localised around $x=0$ and the width of the resulting
kink (or BPS brane) is determined by the parameters in $V(T)$.
These parameters must be tuned such that ${\cal T}_{p-1}$ takes
the correct value. Furthermore, by considering small perturbations
about the solution $T(x,t) = \pm x + f(x,t)$ and linearising the
equations of motion, it is possible to show that if $E$ is finite,
solutions (\dr{kink1}) are stable against small perturbations.

When $q=1/2$, the case we focus on here, the square bracket in
(\dr{derrickT}) is $p+(p-1) \partial_i T \partial^i T$ which can
never vanish for $p \geq 1$. Thus no finite energy static
solutions seem to be permitted.  However, there is formally a way
around this: when $p=1$ (\dr{derrickT}) becomes
\be
0=\int d^p x \frac{V(T)}{ \sqrt{1 +  T'^2 }}
\dle{derrickT2}
\ee
which can vanish if
$$T' \rightarrow \pm \infty
\qquad {\rm or} \qquad T \rightarrow \infty
$$
with $E[T]$ remaining finite.  Let us suppose that the
equations of motion admit a solution with $T' \rightarrow \pm
\infty$ describing a {\it single} kink (below we will find the
conditions on $V(T)$ such that this is the case),
\be
T(x) = \lim_{C \rightarrow 0} \frac{x}{C} = \left\{
\begin{array}{cl}
\infty & x > 0 \\
0 & x=0. \\
-\infty & x < 0
\end{array}
\right.
\dle{SenT}
\ee
This is a typical case of the solutions discussed by Sen \dc{Sen:2003tm}: it
describes an infinitely thin topological kink interpolating
between the two vacuua. Notice that $V=0$ everywhere apart from
when $T=0=x$. Substitution of (\dr{SenT}) into the energy
functional (\dr{Etach}) (taking carefully the $C \rightarrow 0$
limit) gives the the energy of this singular solution to be
\be
E_{\rm Sen} = \int_{-\infty}^{\infty} V(x) dx.
\dle{energythin}
\ee
Sen has argued that such singular kinks are stable, as the BPS
brane should be, and furthermore that their effective action is
exactly the required DBI action \dc{Sen:2003tm}.  Hence such
solutions are of great interest, and would suggest that BPS branes
are infinitely thin.  Once again the parameters of $V(T)$
should be tuned such that $E_{\rm Sen} = {\cal T}_{p-1}$.

The question we ask in this paper is the following: do the
equations of motion for $T$ always admit a single infinitely thin
singular kink solution with energy given by (\dr{energythin})?
 As we will
show, the answer depends sensitively on the shape of the tachyon
potential $V(T)$ for large $T$.  For instance, for $V(T) \sim \exp(-T^a)$
with $a \geq 1$, this solution does not
exist.\footnote{For these potentials, one finds instead an array
of singular kink anti-kinks whose energy is divergent --- see section 3.}
Our approach will be to
try to find a regularised kink solutions from which Sen's singular
limit can then be approached.  This can be
done through compactification.

\section{Regularised kinks  and  examples}

We are looking for static kink-like solutions in which $T$ has a
non-trivial dependence only on one spatial coordinate; $T=T(x)$.
Let as assume that the kink is centered at the origin,
\be
T(0)=0.
\ee
The equations of motion coming from (\dr{actiongen}) with $q=1/2$
have a first integral,
\be
{V(T) \over \sqrt{1+(T')^2}}=V_0,
\dle{v0}
\ee
where $V_0 \geq 0$ is a constant. The energy of this solution is
given by
\be
E=\int dx V(T)\sqrt{1+(T')^2} = \frac{1}{V_0} \int dx V^2(T).
\dle{E}
\ee
Since $T'^2$ is positive, solutions of equation (\dr{v0}) exist in
the region
\be
V(T) \geq V_0 \qquad ({\rm soliton \; solution)}.
\ee
%Writing equation (\dr{v0}) as
%\be
%T'^2 - \frac{1}{\alpha'} \left( \frac{V}{V_0}
%\right)^2=-\frac{1}{\alpha'}
%\ee
%(a `particle' with total energy $-1/\alpha'$
Furthermore, the solutions $T(x)$ are periodic (see figure 1) with a
$V_0$-dependent amplitude which, from (\dr{v0}), diverges as $V_0
\rightarrow 0$. Note that within one period there must be both a
kink and an anti-kink (corresponding to the two points for which
$T(x)=0$), as can clearly be seen in figure 1.  Also, the energy
density of the kinks becomes more and more localised as $V_0
\rightarrow 0$. Below we will study in detail the dependence of
both the period and $E$ on $V_0$.

\begin{figure}%[ht]
\centerline{ \scalebox{0.4}{
\includegraphics{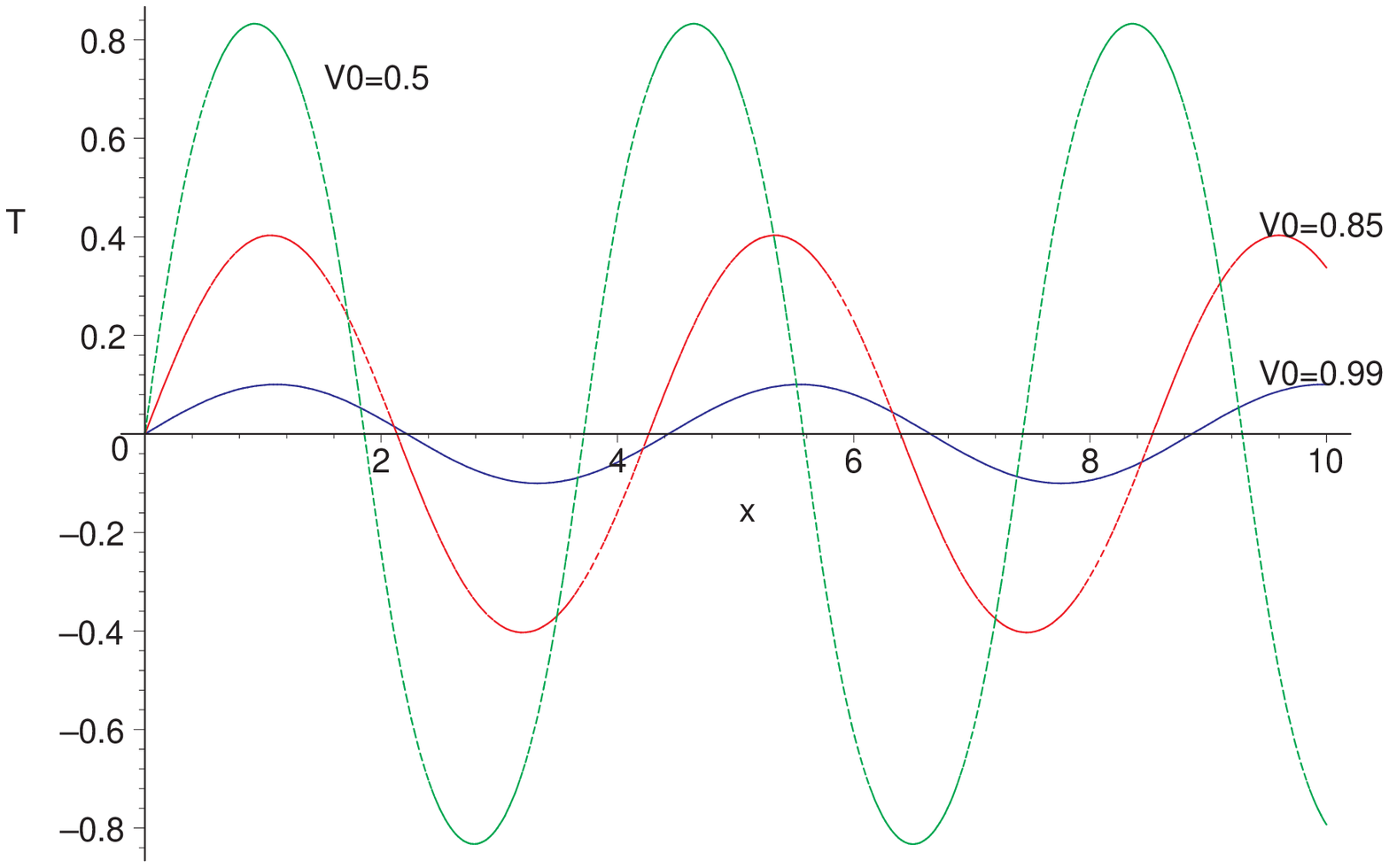}}
\hspace*{1cm}
%\raisebox{0.5cm}{
 \scalebox{0.4}{ \includegraphics{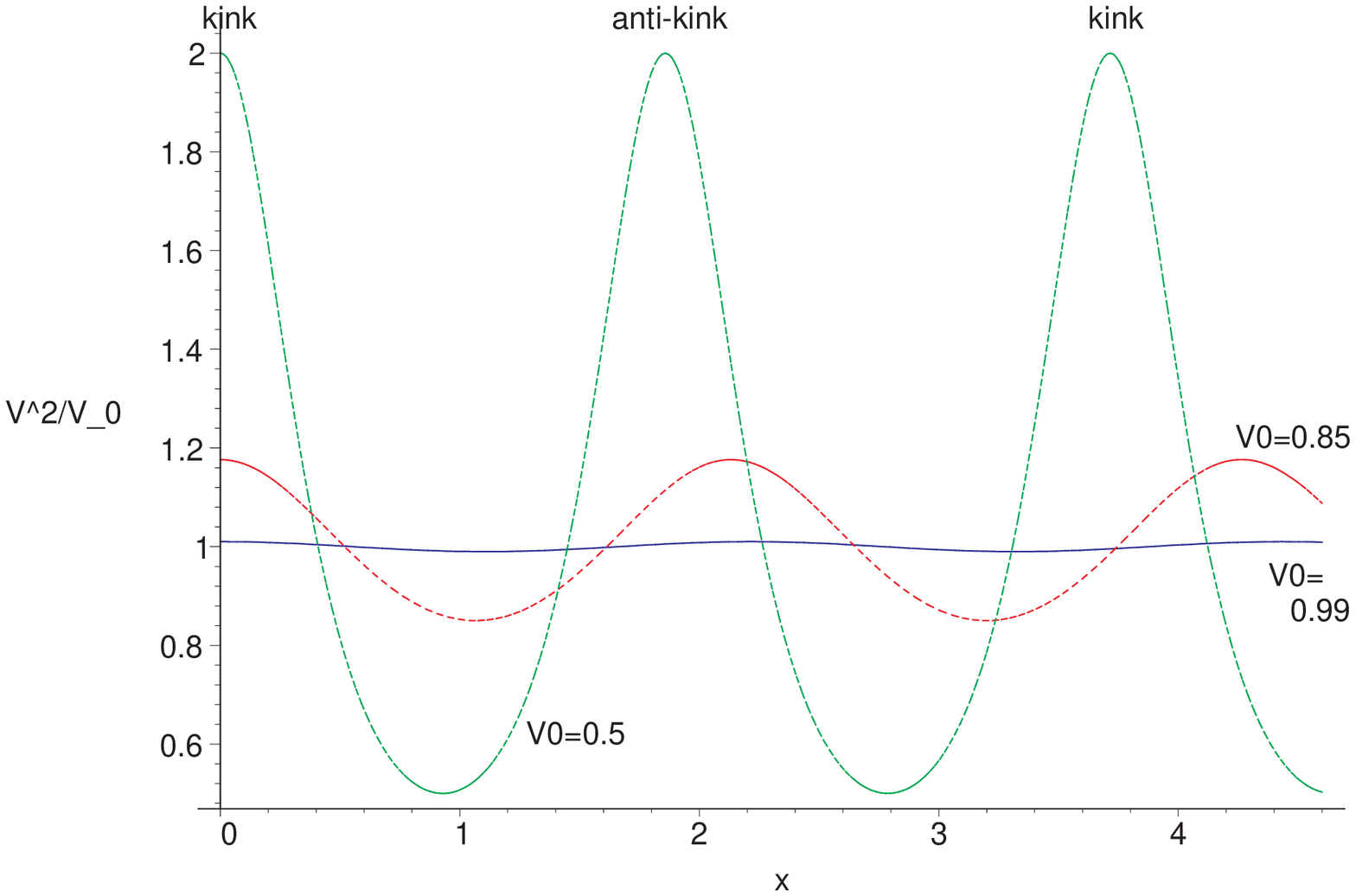}}}
 %}
\caption{LH panel: The solution $T(x)$ of equation (\dr{v0}) for the potential
$V = e^{-T^2}$. As discussed in the text, the solutions are periodic and
their amplitude increases as $V_0$ decreases: the 3 curves correspond to
$V_0 = 0.5$ (green curve), $V_0=0.85$ (red curve), and $V_0=0.99$ (blue 
curve).  Note, as will be discussed later, that the wavelength of the solutions
decreases as $V_0$ decreases.  RH panel: the energy density $V^2/V_0$
corresponding to the 3 curves in the LH panel. The kinks and anti-kinks are 
labeled. Note that as $V_0$ decreases
the energy density in the kinks and anti-kinks have more
localised energy density.}
\dle{fig1}
\end{figure}

In order for $E$ to be finite, one must have $|T| \rightarrow
\infty$ as ${|x| \rightarrow \infty}$. This immediately implies
from (\dr{v0}) that $V_0 = 0$
--- a topological kink.\footnote{Notice that this condition is
identical to (\dr{derrickT2}).}
 Sen's solution \dc{Sen:2003tm} (\dr{SenT}) is such a
singular solution in which $T$ vanishes for only one value
of $x$: in other words Sen's solution is periodic with a divergent
wavelength.  Furthermore in that case $E = 2E_{\rm Sen}$.

We would like to approach the singular ($V_0 \rightarrow 0$) kink(s)
as the limit of regular solutions with $V_0 \neq 0$.  In order to
get regular and finite energy solutions we shall suppose that $x$
is a compact direction of length $2\pi R$, so that $V_0$ can now
be non-zero. Our aim is to see whether the limit
$R\rightarrow\infty$ with $E$ being finite exists. Note that
if this limit exists we expect to get twice (\ref{energythin}) as
the energy of the resulting solution. We can see this from several
points of view: BPS branes have a RR charge and on a circle the
sum of the charges must vanish so all we can get are pairs of
branes and anti-branes. Sen's solution corresponds to the brane
infinitely distant from the anti-brane. Alternatively note
that on a circle the solution is periodic, so if a kink
is present at the origin, where $T'$ is positive, then necessarily
an anti-kink will also be present since as $\int_0^{2 \pi R} T' dx=0$,  $T'$ must
be negative somewhere between $0$ and $2\pi R$. So if we can regularise
Sen's solution by compactifying the $x$ coordinate, the limiting
energy is expected to be twice the value (\ref{energythin}).  Half
of this corresponds to the anti-kink which is sent to infinity, so that
we recover $E_{\rm Sen}$ for the remaining kink.

Since the wavelength of the kink solutions is $V_0$ dependent, not
all values of $V_0$ will be permitted for a given radius $R$. The
dependence of $V_0$ on $R$ can be determined as follows: the
period of the solutions of equation (\ref{v0}) is $4\zeta$ where
\be
\zeta(V_0) =\int_{0}^{T_0}{dT \over
{\sqrt{\left(V\over V_0\right)^2-1}}},
\dle{zeta}
\ee
and $T_0$ is defined by $V(T_0)=V_0$. The radius is thus given by
\be
2 \pi R=4n\zeta,
\dle{ra}
\ee
where $n=1,2,\dots$. The separation between the brane and the
anti-brane is $2\zeta$. The $V_0$-dependent energy of this
solution can be written as
\ba
E(V_0) &= &4n\int_{0}^\zeta dx {V^2(T)\over V_0} = 4n
\int_{0}^{T_0}dT{ V \over{\sqrt{1-\left( V_0\over V\right)^2}}}
\dle{ener}
\\
&\equiv & 4 n{\cal{E}}(V_0) .
\dle{epsdef}
\ea
To summarise, $V_0$ is implicitly determined by equation
(\ref{ra}) in terms of an integer $n$ and the radius $R$. The
energy of the solution depends on $R$ and on $n$. Note
that this
dependence is complicated by the presence of $V_0$ in
(\ref{ener}).

The limit when $V_0\rightarrow 0$ or equivalently $T_0\rightarrow
\infty$ is not transparent for $\zeta$ and $\cal E$. There is, however,
a combination of both for which the limit is known: consider
\be
\Delta(V_0) \equiv {\cal E}-\zeta V_0=\int_{0}^{T_0}
dT\sqrt{(V^2-V_0^2)}.
\dle{DeltaV0}
\ee
Then as $V_0$ goes to zero\footnote{This result is due to the monotone
convergence theorem.}
\be
\Delta(V_0)\rightarrow \int_{0}^{\infty} dT \; V(T) \label{rel},
\ee
which is exactly $E_{\rm Sen}/2$.

Let us examine the limit where the radius goes to infinity. There
are three possibilities depending on the behaviour of $\zeta(V_0)$
as $V_0 \rightarrow 0$:
\begin{enumerate}
\item
$\zeta \rightarrow \infty$ as ${V_0 \rightarrow 0}$.
The kink
anti-kink separation tends to infinity, and as $R \rightarrow
\infty$, equation (\dr{ra}) can be satisfied with $n=1$ --- a
single kink. From equation (\ref{rel}) we see that
the energy goes to $2\int_{-\infty}^{\infty} V(T) dT$
if and only if $\zeta V_0\rightarrow 0$.

\item $\zeta \rightarrow$ const $\neq 0$ as ${V_0 \rightarrow 0}$.
Now the kink anti-kink distance tends to a constant (independently
of $R$) and one can never get an isolated kink.  As $R \rightarrow
\infty$, equation (\dr{ra}) can only be satisfied if $n
\rightarrow \infty$.  In this case (\ref{ener}) shows that that
are no finite energy solutions in the decompactified limit. For a
given finite $R$, $E$ is however finite. \item $\zeta \rightarrow
0$ as $V_0\rightarrow 0$.  For a fixed $R$, the limit
$V_0\rightarrow 0$ now leads to an array of infinitely many
singular kinks and anti-kinks (since their separation is $2
\zeta(0)$).  From (\dr{epsdef}) the energy (even for finite $R$)
is infinite. This is a rather problematic (and not very
meaningful) limit, but we will see that this behaviour occurs for
many potentials considered in the literature; for example $V(T) = v e^{-T^2}$.
\end{enumerate}

As we shall show in the next subsection, the behaviour of
$\zeta(V_0)$ as $V_0 \rightarrow 0$ depends critically on the form
$V(T)$ for large $T$.  Before doing that, we present a very simple
example of case (2) above, and hence a potential for which Sen's
solution can not be recovered.
%We now show, through a simple example, that the solution of (\dr{v0}) does not
%necessarily admit a single singular kink in the limit $V_0 \rightarrow 0$;
%in other words, in that limit $\zeta$ does not necessarily diverge.
%Furthermore, if there is more than one kink, we will see that
%its energy is no longer given by
%(\dr{energythin}).
Let
\be
V(T) = \frac{v}{\cosh \beta T}.
\dle{Vcosh}
\ee
(This potential has also been analysed elsewhere \dc{cosh} and
motivated from string theory calculations in
\dc{Kutasov:2003er}.) On substitution into (\dr{v0}) it is straightforward to
solve for $T(x)$: one finds
\be
T(x) = {\rm arc}\sinh \left[ \frac{1}{\beta} \left(
\frac{v^2}{V_0^2} - 1 \right)^{1/2} \sin (\beta x) \right],
\dle{solncosh}
\ee
which is indeed periodic with a divergent amplitude in the
singular $V_0 \rightarrow 0$. Notice, however, that the 
period is independent of $V_0$:
\be
\zeta(V_0) = \frac{\pi}{2\beta} \; ,\qquad \forall \; V_0.
\ee
Thus as $V_0 \rightarrow 0$, $\zeta$ remains constant leading to
an array of localised kink-anti-kink pairs. Let us now compactify
on the circle of radius $R = 4 n \zeta/2 \pi =n \pi/\beta$, where
$n=1,2,\ldots$ counts the number of kinks. (This limiting
potential has the unusual feature that for a given $R$, $V_0$ need
not take discrete values: all values between $0$ and $v$ are allowed.)
Using (\dr{solncosh}) the energy (\dr{ener}) of the solution is
easily obtained, after a contour integration, to be
\be
E = \frac{4n}{V_0} \int_0^{\frac{\pi}{2\beta}} \frac{v^2}{\cosh^2
(\beta T(x))} dx = \frac{2 \pi v}{\beta} n = 2 \pi R v
\dle{Ecrit}
\ee
i.e.\ independent of $V_0$.  Thus each solution is equally
energetic. In the limit $R \rightarrow \infty$ the energy diverges
and is clearly not given by (\dr{energythin}).

In the next subsection we determine the conditions which $V$ must
satisfy so that in the decompactified limit with a finite energy, $\zeta$ 
diverges.

\subsection{Solitonic solutions on the circle}

From the discussion in the previous section, the dependence of
$\zeta$ on $V_0$ when $V_0$ is small is crucial. In order to
analyse this dependence let us parametrise the general potential
$V(T)$ by
\be
V(T) = \frac{v}{\cosh f(T)}
\ee
where $f(0)=0$. Then, on letting
\be
u = \frac{ \sinh f(T)}{\sinh f(T_0) },
\ee
the period $\zeta(V_0)$ in (\dr{zeta}) becomes
\be
\zeta(V_0) = \int_0^1 du \frac{1}{\sqrt{1-u^2}} \frac{1}{\left|
\frac{df(T(u,V_0))}{dT} \right|}.
\dle{zetanice}
\ee
Notice immediately that if $f(T)=T$ (i.e.\ the potential discussed
in the previous section), we recover the constant $\zeta$.
Clearly the dependence of $\zeta(V_0)$ is determined by $f'$.  The
$V_0$-dependence of $f'(T(u,V_0))$ is through the combination $u
\sqrt{ (v/V_0)^2 - 1 }$.
Thus from (\dr{zetanice}), the behaviour of $\zeta(V_0 \rightarrow
0)$ depends on $f'(T \rightarrow \infty)$. Since $|f'| \sim |V'|/V$
it follows that
\be
\zeta(V_0) \rightarrow \frac{\pi}{2}
\left. \frac{V}{|V'|} \right|_{T_0} \qquad (V_0
\rightarrow 0 \; , \;  T_0 \rightarrow \infty ).
\dle{zeta0}
\ee

Since the potential is assumed
positive let us write it in the form $V=e^{-\sigma(T)}$
then the
condition for $\zeta$ to diverge in the limit $T_0\rightarrow
\infty$ reads
\be
\lim_{T\rightarrow \infty} \sigma'(T)=0.\label{con}
\ee
This is one of our main results.
This condition is clearly not satisfied for the $e^{-aT^2}$
potential. In fact this potential gives rise to $\zeta$ which
vanishes in the $V_0\rightarrow 0$ limit: an example of case 3
discussed in the previous section.  Thus for exponential
potentials, in the limit $V_0 \rightarrow 0$,
\be
V(T \rightarrow \infty) \sim \exp(-T^{a}) \qquad \Rightarrow
\qquad \left\{
\begin{array}{ll}
a < 1, & \zeta \rightarrow \infty  \\
a > 1, & \zeta \rightarrow 0  \\
a = 1, & \zeta \rightarrow K \neq 0
\end{array}
\right.
\dle{conexp}
\ee
For power-law potentials, again in the limit $V_0 \rightarrow 0$:
\be
V(T \rightarrow \infty) \sim \frac{1}{T^{1/\alpha}} \qquad
\Rightarrow  \qquad \left\{ \alpha
> 0, \qquad \zeta \rightarrow \infty \right.
\dle{conpower}
\ee
where the condition on $\alpha$ simply enforces that $V(T)$
vanishes at infinity.

%
% Another way of stating
%condition (\ref{con}) is that the potential should vanish at
%infinity slower than any power of $e^{-T}$. For example
%$e^{-b(T^2)^\alpha}$, with $\alpha <1/2$ and $b$ positive are
%acceptable potentials near infinity.

Before turning to the energy, it is useful to derive two further
results for the period $\zeta(V_0)$.  The first is the value of
$\zeta(V_0=v)$. Let us suppose that $V_0 = v - \epsilon$,
($\epsilon \ll v$), so that the tachyon oscillates around $T=0$
with a very small amplitude. Thus $V(T) = v(1 -
\beta^2 T^2/2) + {\cal{O}}(T^4)$ where $\beta$ is dimensionless, and
the solution of (\dr{v0}) is
\be
T(x) = \frac{1}{\beta^2} \left(1 - \frac{ V_0^2}{v^2} \right) \sin
( x \beta ).
\ee
Hence
\be
\zeta(v) =\frac{\pi}{2 \beta}\qquad {\rm and}  \qquad \left.
\frac{d\zeta}{dV_0} \right|_{V_0 = v} = 0.
\dle{derivszeta}
\ee

Secondly, it will be useful to express $T(V)$ as an integral over
$\zeta(V_0)$.  To do so, note that since
$$ \int_a^b \frac{dx}{\left[ (x-a)(b-x) \right]^{1/2}} = \pi
$$
for arbitrary $a$ and $b>a$,
equation (\dr{zeta}) can be inverted to give
\be
T(V) = \frac{1}{\pi} \int_{V}^{v} dx \,  \frac{\zeta(x)}{(x^2 -
V^2)^{1/2}}
 = \frac{1}{\pi} \int_0^{{\rm arc}\cosh(v/V)}  d \theta \;
\zeta(V \cosh \theta)
\dle{TV1}
\ee
where $x = V \cosh \theta$.  Thus if we specify a particular
behaviour of $\zeta(x)$
we can use (\dr{TV1}) to determine $T(V)$ which can
then be inverted to find $V(T)$. This will be done in subsection
3.2 where it
will be useful to note from (\dr{TV1}) that the vanishing of the
potential as $|T| \rightarrow \infty$ imposes the condition
\be
\int_0^v dV_0 \; \left( \frac{\zeta(V_0)}{V_0} \right)\rightarrow
\infty.
\dle{condint}
\ee

We now ask what conditions $V(T)$ must satisfy such that as $V_0
\rightarrow 0$ and $R \rightarrow \infty$ the energy $E$ in
(\dr{ener}) with $n=1$ reduces to $2E_{\rm Sen}$ in (\ref{energythin}). First
notice from equations (\dr{DeltaV0}),
(\ref{rel}) and (\dr{zeta0}) that ${\cal{E}}(V_0)$ will be
finite if and only if, as $T\rightarrow \infty$,
\be
 {|V'| \over V^2}\rightarrow \infty.
 \dle{epsilon0}
 \ee
This condition should be contrasted with the condition
(\dr{zeta0}) on $\zeta$.  In this case, ${\cal E} \rightarrow \int_0^{\infty} V(T)dT$.
Thus from (\dr{epsilon0}) we conclude
that for exponential potentials
\be
V(T \rightarrow \infty) \sim \exp(-T^{a}) \qquad \Rightarrow
\qquad {\cal{E}} \; {\rm finite} \; \forall \; a > 0
\dle{conexpenergy}
\ee
whereas for power-law potentials
\be
V(T \rightarrow \infty) \sim \frac{1}{T^{1/\alpha}} \qquad
\Rightarrow \qquad {\cal{E}} \; {\rm finite} \; \iff \; \alpha <
1.
\dle{conpowerenergy}
\ee

Now, recalling that $E = 4 n {\cal{E}}$, we can ask whether $E$ is
finite in the decompactified limit $R \rightarrow \infty$ as $V_0
\rightarrow 0$. Combining the conditions (\dr{conexp}) and
(\dr{conexpenergy}) for exponential potentials gives, in these
limits
$$V(T) \sim \exp(-T^{a}) \; \; {\rm as} \; \; R \rightarrow \infty \; \; {\rm and} \; \; V_0 \rightarrow
0:$$
\be
\begin{tabular}{|c|c|}
  \hline
  % after \\: \hline or \cline{col1-col2} \cline{col3-col4} ...
 { $\begin{array}{c}
  {\cal{E}} \rightarrow \int_0^\infty V(T)\, dT\\ \forall \; a >
  0\end{array}$}
    & {$
    \left\{
\begin{array}{lllll}
 a < 1: & \zeta \rightarrow \infty, & n = 1, & E
\rightarrow 2 \int_{-\infty}^{\infty} V(T) \, dT &  \\
 a = 1: & \zeta \rightarrow K,  & n \rightarrow \infty, & E \rightarrow \infty &  \\
a > 1: & \zeta \rightarrow 0,  & n \rightarrow \infty, & E
\rightarrow \infty &
\end{array}
\right.
 $}  \\
%  \hline
  \hline
\end{tabular}
\dle{concexp}
\ee
For power-law potentials, one finds instead
$$V(T) \sim \frac{1}{T^{1/\alpha}} \; \; {\rm as} \; \; R \rightarrow
\infty \; \; {\rm and} \; \; V_0 \rightarrow 0:$$
\be
\begin{tabular}{|c|c|}
  \hline
  % after \\: \hline or \cline{col1-col2} \cline{col3-col4} ...
 { $\begin{array}{c}
  \zeta \rightarrow \infty \; {\rm and} \; n = 1 \\
  \forall \; \alpha
> 0
  \end{array}$}
    & {$
    \left\{
\begin{array}{lll}
\alpha \geq 1, &  E
\rightarrow \infty & \\
\alpha < 1, &  E \rightarrow 2 \int_{-\infty}^{\infty} V(T)
\, dT  & \\
\end{array}
\right.
 $}  \\
%  \hline
  \hline
\end{tabular}
\dle{concpower}
\ee

\subsection{Some examples and solutions for $\zeta(V_0)$}

In this subsection we give the exact $\zeta$ behaviour (focusing
mainly on the $V_0 \rightarrow 0$ limit) for both power-law and
exponential potentials.

First suppose that
\be
\zeta(V_0) \sim V_0^{-\alpha}.
\dle{powerlawzeta}
\ee
Then (\dr{condint}) imposes that
\be
\alpha > 0
\dle{betacon}
\ee
so that $\zeta$ always diverges in the $V_0 \rightarrow 0$ limit.
Note, however, that (\dr{powerlawzeta}) cannot be valid $\forall
\; V_0$: though it can be normalised correctly at $V_0=v$, its
gradient will not vanish there, as required by (\dr{derivszeta}).
Hence let us first consider the large $T$ or small $V_0$
behaviour. What is the potential $V(T)$ corresponding to
(\dr{powerlawzeta})? This can be obtained by a simple substitution
into (\dr{TV1}): for large $T$
\be
T(V) \sim V^{-\alpha} \qquad \Longrightarrow \qquad V(T) \sim
\frac{1}{T^{1/\alpha}}.
\dle{respower}
\ee
Hence we have recovered part of the results of table
(\dr{concpower}): for $V(T) \sim \frac{1}{T^{1/\alpha}}$, one must
have $\alpha > 0$ and furthermore in that case $\zeta$ diverges as
$V_0^{-\alpha}$.  Alternatively, starting from (\dr{respower}) we
could have deduced (\dr{powerlawzeta}) from (\dr{zeta0}).

Furthermore it is useful to notice from (\dr{zetanice}) that if
\be
V(T) = \frac{v}{T^\alpha} \qquad ({\rm i.e} \; \cosh f(T) =
T^{\alpha})
\ee
then
\be
\frac{d \zeta}{dV_0} \leq 0 \qquad \forall \; V_0.
\dle{derivzetapower}
\ee
Thus, as an example, one could take
\be
\zeta(V_0) = \frac{\pi}{2\beta} v^{2\alpha} \frac{1}{[V_0(2v -
V_0)]^\alpha}
\dle{exact}
\ee
which satisfies all the necessary criteria for $\zeta$ (see
(\dr{derivszeta},\dr{condint},\dr{derivzetapower})) and can be
integrated exactly for various values of $\alpha$ (such as
$\alpha=1/2,1,2$). The answer is not particularly illuminating,
though a simple plot shows that it yields a potential $V(T)$ with
vanishing first derivative at $T=0$ and which goes as
$T^{-1/\alpha}$ for large $T$.

Secondly suppose that,
\be
\zeta(V_0) \sim \frac{1}{(- \ln(V_0))^{\frac{a-1}{a}}},
\dle{logzeta}
\ee
so that as $V_0 \rightarrow 0$,
\be a>1 \Rightarrow \zeta \rightarrow 0 \, , \qquad
a<1 \Rightarrow   \zeta \rightarrow \infty \, , \qquad a=1
\Rightarrow K \neq 0.
\dle{zetabehexp}
\ee
Again, note that (\dr{logzeta}) cannot be valid for all $V_0$.  
Substitution of (\dr{logzeta}) into (\dr{TV1})
shows that for large $T$,
\be
T(V) \sim (-\ln V)^{1/a} \qquad \Longrightarrow \qquad V(T) \sim
\exp({-T^{a}}):
\ee
(\dr{logzeta}) generates exponentially decaying potentials.
Thus (\dr{zetabehexp}) confirms our previous results of
(\dr{concexp}).  Finally, as
above, one can deduce from (\dr{zetanice}) that if $f(T) =
T^{a}$ then for all $V_0$,
\be
\frac{d \zeta}{dV_0} \leq 0 \; , \; a < 1 \qquad , \qquad
\frac{d \zeta}{dV_0} \geq 0 \; , \; a > 1. \qquad \qquad (\forall \; V_0)
\dle{derivzetaexp}
\ee

\subsection{Energy for a fixed finite $R$}

Finally, in this subsection we suppose that $R$ is finite and
fixed, and ask which value of $V_0$ is most favoured
energetically. In other words, we calculate $E_R(V_0)$ and ask for
what value of $V_0$ it is minimal.

Clearly when $V_0=v$, $E_R(v)=2 \pi R v$ since the field $T$ is
constant and vanishing.  Furthermore we have seen in (\dr{Ecrit})
that for the limiting potential $V = v/\cosh(\beta T)$ the energy
is $V_0$ dependent and also given by $E_R(V_0)= 2 \pi R v$.  In
general, though, how does $E_R(V_0)$ behave depending on whether
$\zeta$ diverges or tends to zero as $V_0$ tends to zero?  To see
that, it will be useful to note from (\dr{epsdef}) and (\dr{DeltaV0}) that
\be
E_R(V_0) = 4 n {\cal{E}} = 2 \pi R \left[ \frac{1}{\zeta(V_0)}
\int_0^{T_0} (V^2 - V_0^2)^{1/2} dT + V_0 \right]
\ee
and hence
\be
\frac{d E_R(V_0) }{d V_0} = - 2 \pi R
\frac{1}{\zeta^2} \frac{d \zeta}{d V_0} \int_0^{T_0} (V^2 -
V_0^2)^{1/2} dT.
\dle{derivE}
\ee

Above we deduced that if $\zeta \rightarrow \infty$ as $V_0
\rightarrow 0$ (either with an expotential potential
(\dr{derivzetaexp}) or a power-law one (\dr{derivzetapower})) then
$d\zeta/dV_0 \leq 0$ so that $d E_R(V_0)/d V_0 \geq 0$.
Hence $E_R(V_0)$ is minimal when $V_0$ is
takes its smallest allowed value for that $R$: in that case there
is only 1 kink and anti-kink on the circle ($n=1$), and they have
very well defined profiles. In the infinite $R$ limit, the
distance between the kink and anti-kink tends to infinity and
$E_R$ reduces to $2E_{\rm Sen}$. The behaviour of $\zeta$ and $E_R$
is shown schematically in figure 2.

\begin{figure}%[ht]
\centerline{ \scalebox{0.4}{
\includegraphics{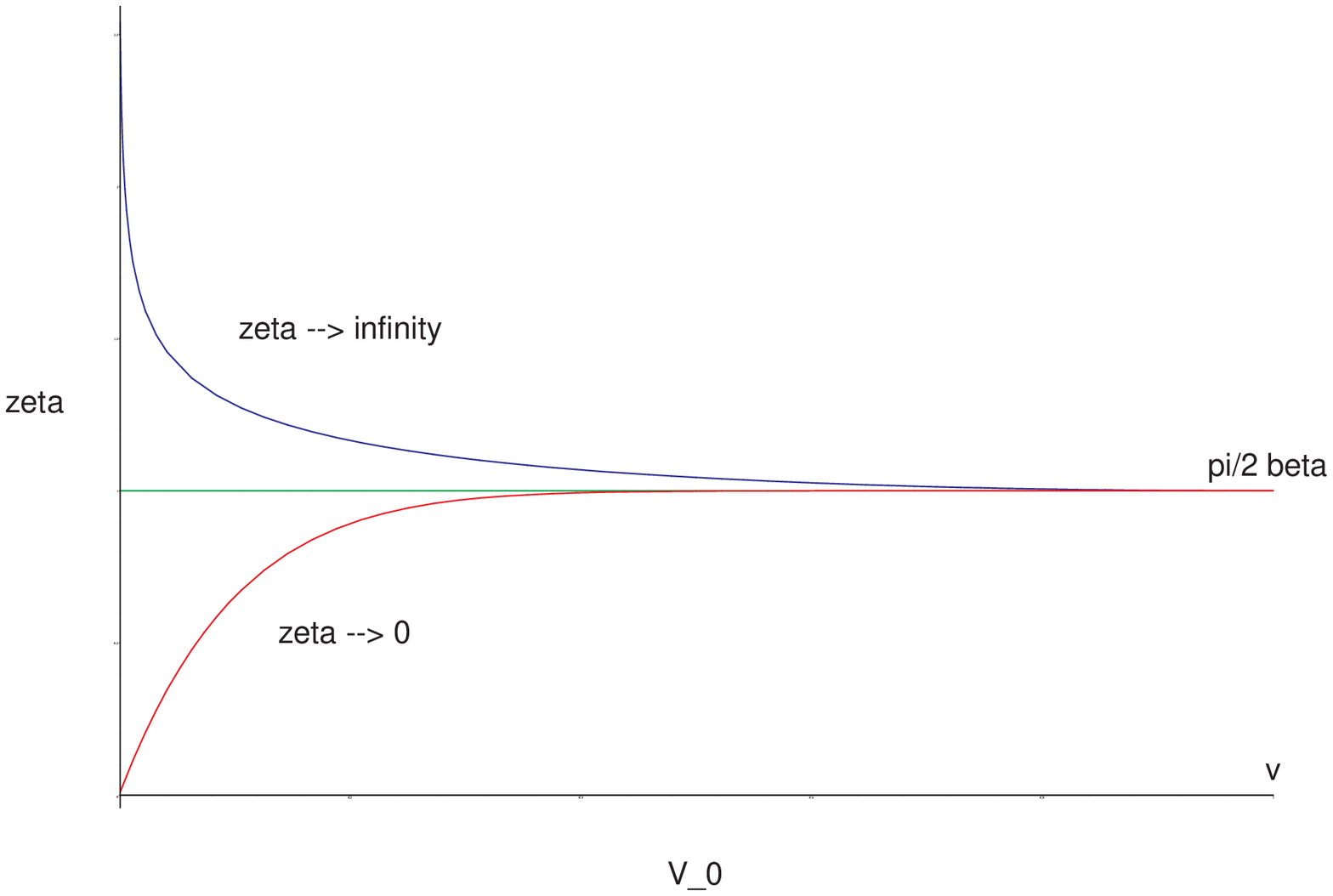}}
\hspace*{1cm}
%\raisebox{0.5cm}{
 \scalebox{0.4}{ \includegraphics{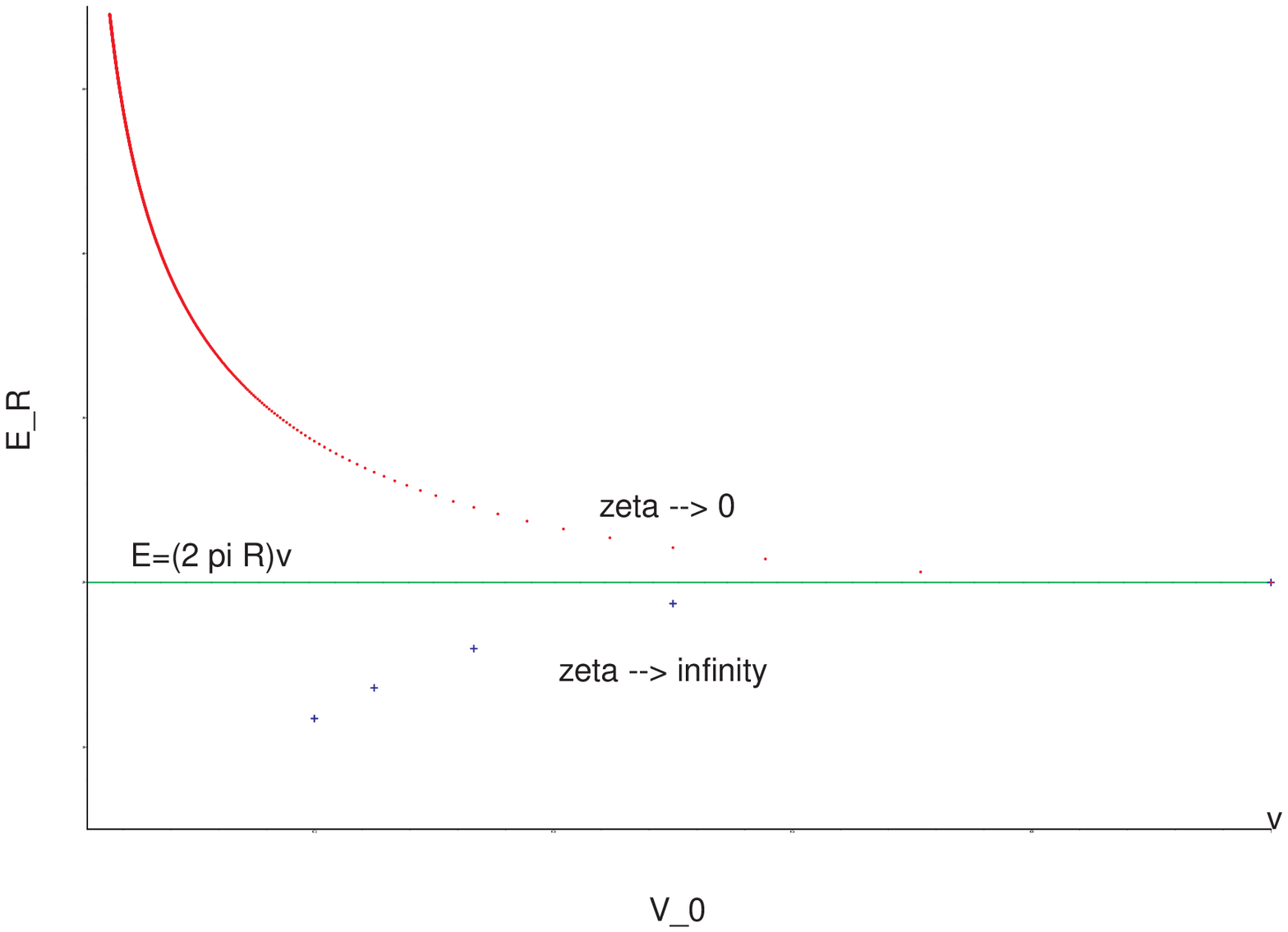}}}
 %}
\caption{Schematic plots of $\zeta$ and $E_R$ as a function of
$V_0$.  In the RH panel, the values of $V_0$ are discrete 
due to the finite size $2\pi R$ of the circle: the exception is
the horizontal line which, in both panels, corresponds to the
potential $V=v/\cosh(\beta T)$.  When $V_0 \rightarrow 0$ and
$\zeta \rightarrow 0$ (red curves in both figures), the energy
diverges even for finite $R$ as seen in the RH panel: the lowest
energy state is the one with constant vanishing $T$ and hence no
kinks, as discussed in the text. Such a situation occurs for
potentials of the form $V \sim e^{-T^a}$, $(a>1)$.  If $\zeta
\rightarrow \infty$ as $V_0 \rightarrow 0$ (blue curves in both
figures), the lowest energy state has the lowest value of $V_0$
allowed for that $R$ and corresponds to a state with one kink and
anti-kink. In the limit $R \rightarrow \infty$, $V_0 \rightarrow
0$ its energy tends to $E_{\rm Sen}.$}
\dle{fig2}
\end{figure}

Now consider the case in which $\zeta \rightarrow 0$ as $V_0
\rightarrow 0$.  As discussed above, this occurs for potentials of
the form
\be
V = v/\cosh(T^a) \qquad a>1,
\dle{eek}
\ee
for which $d\zeta/dV_0 \geq 0$ for all $V_0$. Thus from
(\dr{derivE}), $d E_R(V_0)/ d V_0 \leq 0$.
Furthermore, as $V_0 \rightarrow 0$, the distance between kinks
and antikinks tends to zero so that $n \rightarrow \infty$ and
hence $E_R(V_0 \rightarrow 0) \rightarrow \infty$. This behaviour
of $E_R$ is shown schematically in figure 2. Thus we conclude that
the minimal energy solution is the one in which $V_0 = v$ so that
$T$ is constant and there are no kinks!  
%Thus for exponential
%potentials of the form (\dr{eek}) the least energy solution
%is one with no kinks!

As we will now see, however, all these solutions for finite $R$ 
are in fact unstable since their spectrum of perturbations 
has tachyonic modes.

%The only possibility in which kink-like solutions could perhaps
%exist for the popular potentials such as $V \sim \exp(-T^2)$ is if
%$E_R(V_0)$ could develop a minimum for a value of $V_0 = V_0^{\rm
%min} < v$. Hence $\zeta$ would have to have a local maximum there,
%and this might occur for potentials of the type $V(T) = T^\alpha
%e^{-T^a}$ with $\alpha > 0$.  (The reasons is that for small $T$
%($V_0 \lsim v$) the power-law dominates leading to $\zeta' > 0$,
%and for large $T$ the expotential dominates, $\zeta' < 0$.)
%However, one should note that in any case $V_0^{\rm min} \lsim v$
%so that the resulting kinks will be very flat (with a very ill
%defined profile). 

\section{Perturbation analysis of periodic solitons on the circle}

We now analyse the linear stability of the periodic solitons on
the circle of radius $R$, also taking the $V_0 \rightarrow 0$ and
$R \rightarrow \infty$ limits at the end.  On the circle, the
solution contains an array of $n$ brane-antibranes, and hence we
expect that the spectrum contains $2n$ tachyonic excitations in
one to one correspondence with the annihilation channels of each
brane-antibrane pair. To see this, we expand the Lagrangian to
second order around a periodic soliton, ${\cal T}$, and write
$$
T={\cal T}(x)+\delta T(x,t)
$$
where, for simplicity, we neglect the dependence of $\delta T$ on
the $p-1$ remaining directions. Then to second order
\ba
{\cal S}^{(2)} &=& -\frac{V_0}{2} \int_0^{2 \pi R} dx \int  dt \;
\left[ \frac{V_0^2}{V^2} \left( \frac{d (\delta T)}{dx} \right)^2 -
\left( \frac{d (\delta T)}{dt} \right)^2 +
\frac{V V'}{V_0^2} (\delta T)^2 \right.
\nn
\\
& & \qquad \qquad \qquad \qquad \qquad \qquad \qquad \qquad \left.
+ 2 \frac{V'}{V} \sqrt{ \frac{V^2}{V_0^2} - 1} \delta T \frac{d
(\delta T)}{dx} \right]
\dle{S2}
\\
&= & - \frac{V_0}{2} \int_0^{2 \pi R} dx \int  dt \; \left[
a({\cal T}) \left( \frac{d (\delta T)}{dx} \right)^2 + b({\cal T})
(\delta T)^2 \right].
\dle{S2bis}
\ea
Here, in going from (\dr{S2}) to (\dr{S2bis}) we have integrated
by parts; $V = V({\cal T}(x))$; and
\be
a = \frac{V_0^2}{V^2} \qquad , \qquad b= - \omega^2 +  \frac{V
V'}{V_0^2} - \frac{d}{dx} \left( \frac{V'}{V} \sqrt{
\frac{V^2}{V_0^2} - 1} \right)
\ee
where we have supposed that
$$
\delta T (x,t) = \cos(\omega t + \epsilon) \delta T(x).
$$
The equation of motion is thus
$$
\frac{d}{dx} \left[ a \frac{d \delta T}{dx} \right]  - b \delta T
= 0,
$$
from which one can verify that
\be
{\delta T} = \frac{d {\cal T}}{dx}
\dle{zeromode}
\ee
is indeed the zero mode corresponding to translation invariance.
In order to obtain a Schr\"odinger equation with the standard
$L^2$ scalar product, the coefficients of the spatial and
time derivatives must be equal (up to a sign) and $x$-independent.
The changes of function and variable
\be
dx = \frac{V_0}{V} dz \qquad {\rm and} \qquad
\delta T = \sqrt{\frac{V}{V_0}} \phi
\dle{changes}
\ee
yield the Schr\"odinger equation
\be
-\frac{d^2\phi}{dz^2}+ \left(\frac{1}{\phi_0} \frac{d^2 \phi_0}{dz^2} \right)
\phi=\omega^2\phi
\dle{Schr}
\ee
where, from (\dr{zeromode}), $\phi_0 \propto \frac{d{\cal T}}{dx}
\sqrt{\frac{V_0}{V}}$. Imposing that $\int \phi_0^2 dz
% = \int_0^{2 \pi R} V \phi_0^2/V_0
= 1$ yields the normalised zero mode to be
\be
\phi_0 = \frac{\sqrt{V_0}}{\sqrt{E - 2 \pi R V_0}} \left(
\frac{d{\cal T}}{dx} \sqrt{\frac{V_0}{V}} \right).
\dle{phizeronorm}
\ee

Notice that $\phi_0$ vanishes $2n$ times at the midpoints between
the kinks and anti-kinks, whilst the potential term in (\dr{Schr})
is a smooth function for all $z$. Hence we can apply
Sturm-Liouville theory for periodic boundary conditions with a
non-singular potential --- this guarantees that the spectrum
$\omega^2$ is discrete, and labeled by the number of zeros of the
eigenfunctions. Since the zero mode possesses $2n$ zeros, we
can deduce that there exists $2n$ eigenfunctions
with negative $\omega^2$. These $2n$ eigenfunctions will be localized around the 
$2n$ zeros of $\phi_0$. For any $R$ and non-zero $V_0$ the kink anti-kink array is unstable.
% that we identify with the tachyonic
%excitations of the array of brane-antibranes.

Let us examine the $R\to\infty$ solutions whose only finite energy
possibility is when $\zeta \to\infty$.
One must check how the distance
between the zeros of $\phi_0$ depends on $z$ which (rather than
$x$) is the relevant variable for the Schr\"odinger equation.
Furthermore, how does this distance depend on the behaviour of
$\zeta$ as $V_0 \rightarrow 0$?  Using (\dr{changes}), the
distance between zeros is $2z_0$ where
\be
z_0(V_0) = \int_0^{\zeta(V_0)} \frac{V}{V_0} dx = \int_0^{T_0}
\frac{V}{\sqrt{V^2 - V_0^2}} dT
\dle{z0}
\ee
so that from the definition of $\zeta$,
\be
z_0 - \zeta = \int_0^{T_0} \sqrt{\frac{V-V_0}{V+V_0}} dT \; \;
\longrightarrow \infty \; \; {\rm as} \; V_0 \rightarrow 0
\ee
(again we have used the monotone convergence theorem). 
Thus, independently of the behaviour of $\zeta$, $z_0$ always
diverges as $V_0 \rightarrow 0$ and $\phi_0 \sim
\sqrt{{V(z)}/{2E_{\rm Sen}}}$: the tachyonic modes drop out of
the spectrum in that limit.

\section{Conclusion}

We have shown that a compact dimension of radius $R$ allows the existence of time-independent and
finite energy solutions of the tachyon equations derived from the DBI effective action.
These solutions represent an array of kinks and anti-kinks and are unstable due to tachyonic fluctuations.  Stable solutions
exist in the limit of infinite separation between the kink and anti-kink. 
For this to be occur, three conditions must be satisfied: {\it i)} $ \frac{V'}{V} \rightarrow_{|T| \rightarrow \infty} 0,
$ {\it ii)} 
$\frac{V'}{V^2} \rightarrow_{|T| \rightarrow \infty} \infty,$ {\it iii})$\int_0^{\infty} V(T) dT < \infty$.
The first guarantees that in the non-compact limit, the kink and anti-kink
are infinitely separated.  The remaining ones guarantee the finiteness of the energy.

\section*{Acknowledgements}

D.A.S.\ would like to thank B.Carter and T.Vachaspati for many useful discussions, and the Flora Stone Mather 
Foundation and Physics department of Case Western Reserve University for support and hospitality 
whilst this work was in progress.

\end{document}